\documentclass[aps,prl,twocolumn, preprintnumbers, floatfix, superscriptaddress,pacs]{revtex4}

\usepackage[dvips]{graphics}
\usepackage{color}
\usepackage{amsmath,amsfonts,bm}
\usepackage{epsfig,bm}
\usepackage{graphicx,comment}
\begin{document}

\preprint{JLAB-THY-07-624}

\affiliation{Thomas Jefferson National Accelerator Facility,
             Newport News, VA 23606, USA}
\affiliation{Physics Department, Louisiana State University,
             Baton Rouge, LA 70803, USA}
\affiliation{Physics Department, Old Dominion University, Norfolk,
             VA 23529, USA}
\affiliation{Laboratory of Theoretical Physics, JINR, Dubna, Russian
             Federation}
\title{Form Factors and Wave Functions of Vector Mesons in Holographic QCD}
\author{Hovhannes~R.~Grigoryan}
\affiliation{Thomas Jefferson National Accelerator Facility,
             Newport News, VA 23606, USA}
\affiliation{Physics Department, Louisiana State University,
             Baton Rouge, LA 70803, USA}
\affiliation{Laboratory of Theoretical Physics, JINR, Dubna, Russian
             Federation}
\author{Anatoly~V.~Radyushkin}
\affiliation{Thomas Jefferson National Accelerator Facility,
              Newport News, VA 23606, USA}
\affiliation{Physics Department, Old Dominion University, Norfolk,
             VA 23529, USA}
\affiliation{Laboratory of Theoretical Physics, JINR, Dubna, Russian
             Federation}

\newcommand\sect[1]{\textbf{\emph{#1}} --}

\begin{abstract}
Within the framework of a holographic dual model of QCD, we develop
a  formalism for calculating form factors of vector mesons. We show
that the holographic bound states can be described not only in terms of
eigenfunctions of the equation of motion, but also in terms of
conjugate wave functions that are close analogues of
quantum-mechanical bound state wave functions. We derive a
generalized VMD representation for form factors, and find a very
specific VMD pattern, in which form factors are essentially given by
contributions due to the  first two bound states in the
$Q^2$-channel. We calculate  electric radius of the $\rho$-meson,
finding the value $\langle r_\rho^2 \rangle _C = 0.53 \, {\rm
fm}^2$.

\end{abstract}


\maketitle

\sect{Introduction.} The AdS/CFT correspondence
\cite{Maldacena:1997re} conjectures  equivalence of gravity theory
on the Anti de Sitter space AdS$_5$ and a strongly coupled
four-dimensional (4D) conformal field theory (CFT).  The
correspondence states that for every CFT operator ${\cal O}(x)$
there is a corresponding bulk field $\Phi (x,z)$ uniquely determined
by  the boundary condition (b.c.) $\Phi(x,z=0)$ at the ultraviolet
(UV) 4D boundary of AdS space ($x$ denotes the 4D  coordinates and
$z$ stands for the fifth extra dimension). The addition of an
infrared (IR) brane at $z=z_0$ breaks   conformal invariance in the
IR region, and allows one to have both particles and $S$-matrix
elements.  Due to the holographic equivalence between the broken CFT
and the gravitational picture, the two theories  have identical
spectra and identical $S$-matrix elements
\cite{Arkani-Hamed:2000ds}. In particular, the Kaluza-Klein modes on
the gravity side can be interpreted as bound  states in the 4D
theory. The next  conjecture  is  that the AdS/CFT correspondence
can be extended to assert  that any 5D gravity theory on  AdS$_5$ is
holographically dual to some strongly coupled, large-$N_c$ 4D CFT
(see,  e.g., \cite{Arkani-Hamed:2000ds}). The goal of holographic
models of quantum    chromodynamics (QCD)  is to find such a gravity
theory for which the dual theory is as close to QCD as possible.

Holographic duals of QCD based on the AdS/CFT correspondence have
been applied  recently to hadronic physics (see, e.g.,
\cite{Polchinski:2002jw,Brodsky:2003px,Sakai:2004cn,Erlich:2005qh,Erlich:2006hq,DaRold:2005zs,
Ghoroku:2005vt,Hirn:2005nr,Brodsky:2006uq,Karch:2006pv,Csaki:2006ji,Hambye:2005up}).  These
models are able to incorporate  essential  properties of QCD such as
confinement and chiral symmetry breaking, and have demonstrated in
many cases success in determination of static hadronic properties,
such as resonance masses, decay constants, chiral coefficients, etc.
Dynamic properties (form factors) have   been studied originally
within the holographic   approach of Ref. ~\cite{Polchinski:2002jw},
and the connection between AdS/QCD approach of
Refs.~\cite{Polchinski:2002jw,Brodsky:2003px} and the usual
light-cone formalism  for  hadronic form factors was proposed in
\cite{Brodsky:2006uq}  and  discussed in \cite{Radyushkin:2006iz}.
The calculation   of     form   factors 
of  scalar   and   vector hadrons within the 
approach of Ref.~\cite{Polchinski:2002jw} 
 was   performed   in   Refs.~\cite{Hong:2004sa,Hong:2005np},
and   applied to   study   the     universality  of   the 
$\rho$-meson   couplings to   other   hadrons. 
The expressions  for  hadronic  form factors given  
in    Refs.~\cite{Polchinski:2002jw,Brodsky:2006uq,Hong:2004sa}
have   an   expected    form   of $z$-integral 
containing the  product of    two hadronic   wave    functions
and a  function  describing    the  probing  current.
However,  the hadronic     functions used in Ref.~\cite{Brodsky:2006uq} 
strongly  differ from those in Refs.~\cite{Polchinski:2002jw,Hong:2004sa}.
The   latter give  meson    coupling constants through
their derivatives  at $z=0$  and    satisfy Neumann 
b.c.  at   the   IR boundary $z=z_0$,   while the   functions
used in Ref.~\cite{Brodsky:2006uq}   satisfy Dirichlet
b.c.   at  $z=z_0$,  and are proportional  (after extraction 
of   the overall $z^2$ factor) to  the 
meson   coupling constants $f_n$   at    the   origin.
In   these   respects  they  are    analogous to   the   bound
state wave   functions in   quantum    mechanics,
which   makes    possible their      interpretation
in   terms   of light-cone    variables  
proposed in   Ref.~\cite{Brodsky:2006uq}.

The aim of this letter is to study form factors and wave functions
of vector   mesons   within the framework of   the holographic QCD
model described in
Refs.~\cite{Erlich:2006hq,Erlich:2005qh,DaRold:2005zs} (which will
be referred to as H-model). To this end, we consider a 5D dual of
the simplest $N_f=2$  version of QCD to be a  Yang-Mills theory with
the $SU(2)$ gauge group in the background of sliced AdS space, i.e.,
the 4D global $SU(2)$  isotopic symmetry of $N_f=2$ QCD is promoted
to a 5D gauge symmetry in the bulk. Note, that the AdS/QCD
correspondence  does not  refer  explicitly to quark and gluon
degrees of freedom. Rather, one deals  with the bound states of QCD
which appear as infinitely narrow resonances. The counterparts  in
the correspondence relation are  the vector current $J^a_{\mu}(x)$
with conformal dimension $ \Delta = 3 $  (in QCD, it   may be
visualized  as $\bar{q}(x) t^a \gamma_{\mu}q(x) $ ),   and the 5D
gauge field $ A^a_{\mu}(x,z) $.

We start with recalling the   basic elements of the analysis of
two-point functions $\langle JJ \rangle$ given in
\mbox{Refs.~\cite{Erlich:2005qh,Erlich:2006hq},}   and introduce a convenient
representation for  the \mbox{$A$-field}  bulk-to-boundary
propagator ${\cal  V}(p,z)$ based on the Kneser-Sommerfeld formula
\cite{Kneser}  that gives ${\cal  V}(p,z)$ as an  expansion over
bound state   poles with  the $z$-dependence of each pole
contribution given by ``$\psi$ wave functions'', that are
eigenfunctions of the 5D equation of motion with Neumann b.c. at the
IR boundary. Then we study the three-point function $\langle JJJ
\rangle$ and   obtain  expression for transition form factors that
involves  $\psi$ wave functions and the nonnormalizable mode factor
${\cal J}(Q,z)$. We   write the latter as a sum   over all bound
states in   the channel of electromagnetic current, which gives an
analogue of   generalized vector meson dominance (VMD)
representation for hadronic  form factors.  
As   the next step, we
introduce ``$\phi$ wave functions''  that strongly resemble wave
functions of  bound states in  quantum   mechanics (they satisfy
Dirichlet b.c. at $z=z_0$,  and their values at $z=0$ give  bound
state couplings $g_5 f_n/M_n$, 
 i.e., they   have    the  properties necessary  for 
the light-cone interpretation of   AdS/QCD results 
proposed  in   Ref.~\cite{Brodsky:2006uq}). 
 We rewrite form   factors in terms  of
$\phi$ functions, formulate   predictions for $\rho$-meson form
factors,  and  analyze these predictions in the   regions of small
and  large $Q^2$.

The $\rho$-meson electric radius  is calculated,
and it is also shown that H-model predicts a peculiar VMD pattern
when two (rather   than just one) lowest bound states in the
$Q^2$-channel  play the dominant role while contributions from
higher states can be neglected. 
 This double-resonance dominance  
is   established  both 
for   the $\rho$-meson form   factor $F(Q^2)$ given   by   the   overlap 
of   the $\psi$-wave  function (here  we confirm   the   results 
obtained   in Ref.~\cite{Hong:2004sa}  for 
the   $\rho$-meson  form   factor    considered there) 
and for the form   factor ${\cal F} (Q^2)$  given   by   the   overlap 
of   the $\phi$-wave  function.
 Finally, we summarize our results.


\sect{Two-Point Function.} Our goal is to analyze form factors of
vector mesons within the framework of the holographic model of QCD
based on AdS/QCD correspondence. As a  4D operator on the QCD side,
we  take  the vector current $J^a_{\mu}(x) =
\bar{q}(x)\gamma_{\mu} t^{a}q(x) $, to which corresponds a bulk
gauge field $ A^a_M(x, z) $ whose boundary value is the source for $
J^a_{\mu}(x) $. We follow the conventions of the H-model
\cite{Erlich:2006hq}, with the  bulk fields in the background of the
sliced AdS$_5$ metric
\begin{equation}
ds^2 = \frac{1}{z^2}\left(\eta_{\mu \nu}dx^{\mu}dx^{\nu} -
dz^2\right), {\qquad} 0 \leq z \leq z_0 \ ,
\end{equation}
where $ \eta_{\mu \nu} = \textrm{Diag}\left(1,-1,-1,-1\right) $, and
$ z_0 \sim 1/\Lambda_{QCD}$ is the imposed IR scale. The 5D gauge
action in AdS$_5$ space, corresponding to $ A^a_{M}(x,z) $, is
\begin{align}
S_{\rm AdS} = - \frac{1}{4g_5^2}\int d^4x~dz~\sqrt{g}~{\rm
Tr}\left(F_{MN}F^{MN}\right) \ ,
\end{align}
where $ F_{MN} = \partial_{M}A_{N} - \partial_{N}A_{M} - i
[A_{M},A_{N}] $, $ A_{M} = t^a A^a_M $, ($t^a  \in SU(2),  a=1,2,3 $)  
 and $ M, N = 0, 1, 2, 3, z$.  Since the vector field $
A^a_M(x, z) $ is taken to be non-Abelian, the 3-point function of
these fields in the lowest approximation can be extracted directly
from the Lagrangian.

Before calculating the 3-point function, we recall some properties
of the 2-point function  discussed in \cite{Erlich:2006hq}. Consider
the sliced AdS space with an IR boundary at $z = z_0$ and UV cutoff
at $z = \epsilon$ (taken to be zero at the end of the calculations).
In order to calculate the current-current correlator (or 2-point
function) using the AdS/CFT correspondence, one should solve
equations of motion, requiring the solution at the UV boundary ($z =
0$) to coincide with the 4D source of the vector current, calculate
5D action on this solution and then vary the action (twice) with
respect to the boundary source. The task is simplified when the $A_z
= 0$ gauge is imposed, and the gauge field is Fourier-transformed in
4D, $A_{\mu}(x,z) \Rightarrow \tilde A_{\mu}(p,z)$. Then
\begin{equation}
\tilde A_{\mu}(p,z) = \tilde{A}_{\mu}(p)\, \frac{V(p, z)}{V(p,
\epsilon)} \ ,
\end{equation}
where $ \tilde{A}_{\mu}(p) $ is the Fourier-transformed current
source, and the 5D gauge field $V(p,z)$ is the so-called
bulk-to-boundary propagator obeying
\begin{equation}
\label{Vequation} z\partial_z \left(\frac{1}{z} \,  \partial_z
V(p,z)\right) + p^2 V(p,z) = 0 \ .
\end{equation}
The UV b.c. $ \tilde A_{\mu}(p,\epsilon) = \tilde{A}_{\mu}(p) $ is
satisfied by construction. At the IR boundary (when \mbox{$ z =
z_0$}), we  follow  Ref.~\cite{Erlich:2006hq} 
 (see also Ref.~\cite{Hong:2004sa})  
and  choose the Neumann b.c.
\mbox{$ \partial_z V (p, z_0) = 0 $}  which corresponds to the gauge invariant
condition \mbox{$ F_{\mu z}(x, z_0) = 0 $}. Evaluating the bilinear
term of the action on this solution leaves only the UV surface term
\begin{align}\label{S2}
S_{\rm AdS}^{(2)} = -\frac{1}{2g^2_5}\int \frac{d^4p}{(2 \pi)^4}
\tilde{A}^{\mu}(p)\tilde{A}_{\mu}(p)\left[
\frac{1}{z}\frac{\partial_z
V(p,z)}{V(p,\epsilon)}\right]_{z=\epsilon} .
\end{align}
The 2-point function of vector currents is defined by
\begin{equation}
\label{pimn} \int~d^4x \ e^{i p \cdot x}\langle
J^a_{\mu}(x)J^b_{\nu}(0) \rangle = \delta^{ab} \,
\Pi_{\mu\nu}(p)\Sigma(p^2) \ ,
\end{equation}
where
$ \Pi_{\mu\nu}(p) \equiv \left( \eta_{\mu \nu} -
{p_{\mu}p_{\nu}}/{p^2} \right) $
is  the  transverse  projector.  Varying the action (\ref{S2}) with
respect to the boundary source   produces
\begin{align}
\label{sigmap2}
\Sigma(p^2) = \left.
-\frac{1}{g^2_5}\left(\frac{1}{z}\frac{\partial_z
V(p,z)}{V(p,\epsilon)}\right)\right |_{z = \epsilon \rightarrow 0} \
.
\end{align}
(To get the tensor structure of (\ref{pimn}) by a ``na{\"\i}ve''
variation, one  should  change $A^\mu A_\mu \to  A^\mu
\Pi_{\mu\nu}(p)  A^\nu $ in  Eq.~(\ref{S2})).

 It   is well   known 
(see, e.g., \cite{Polchinski:2002jw,Hong:2004sa})  that 
two linearly independent solutions of Eq.~(\ref{Vequation}) are given
by the Bessel functions $z J_1(Pz)$ and $z Y_1(Pz)$, where $P \equiv
\sqrt{p^2}$. Taking Neumann b.c. for $  V(p,z)$, one obtains
\begin{equation}
V(p,z) = Pz\biggl [ Y_0(Pz_0)J_1(Pz) - J_0(Pz_0)Y_1(Pz) \biggr ] \ ,
\end{equation}
and, hence,
\begin{align}
\Sigma (p^2)& =  \frac{\pi p^2 }{2g^2_5} \left [ Y_0(Pz) - J_0(Pz)
 \frac{Y_0(Pz_0)}{J_0(Pz_0)}\right ]_{z = \epsilon \rightarrow 0} \  .
\end{align}
This expression is singular as $\epsilon \to 0$:
\begin{align}
\Sigma (p^2) = \frac{1}{2g^2_5} \,  p^2 \ln (p^2 \epsilon^2) +
\ldots   \    .
\end{align}
By matching  to QCD result  for $ J^a_{\mu} =
\bar{q}\gamma_{\mu} t^a q $ currents  one finds $ g^2_5 =
12 \pi^2/N_c $ (cf. \cite{Erlich:2005qh}).

The two-point function $\Sigma (p^2)$ has poles when the denominator
function $J_0(Pz_0) $ has zeros, i.e., when $Pz_0$ coincides with
one of the roots $\gamma_{0,n}$ of the Bessel function $J_0 (x)$.
These poles can be explicitly displayed by incorporating the
Kneser-Sommerfeld expansion \cite{Kneser}
\begin{align}
\label{KS} &\frac{Y_0(Pz_0)J_0(Pz) - J_0(Pz_0)Y_0(Pz)}{J_0(Pz_0)}
\nonumber \\[10pt]  &=
-\frac4\pi  \, \sum_{n = 1}^{\infty} \frac{J_0(\gamma_{0,n}z/{z_0})}
{[J_1(\gamma_{0,n})]^2(P^2z^2_0 - \gamma^2_{0,n})} \ ,
\end{align}
valid for $ z \leq z_0 $  (the case we are interested in). Taking
formally  $ z = 0 $  gives  a logarithmically divergent series
reflecting the $\ln \epsilon$ singularity of  the  $z=\epsilon$
expression. Thus,  some  kind of  regularization for this divergency
of the sum is implied. Under this assumption,
\begin{align}\label{vecor}
\Sigma(p^2) &= \frac{2 p^2}{g^2_5 z^2_0}\sum_{n = 1}^{\infty}
\frac{[J_1(\gamma_{0,n})]^{-2}}{p^2 - M^2_{n}} \ ,
\end{align}
where $ M_{n} = \gamma_{0,n}/z_0 $. Hence,  the 2-point correlator
of the H-model has poles when   $P$   coincides   with  one  of
$M_{n}$'s.  Given   that   the   residues  of  all  these poles are
positive,  the poles    may   be  interpreted as bound  states  with
$M_{n}$'s  being their masses. The coupling $f^2_{n}$  with   which
a particular  resonance contributes   to   the total   sum  is
determined by
\begin{align}
f^2_{n} = \lim_{p^2 \rightarrow M^2_{n}}\left \{(p^2 - M^2_{n}) \,
\Sigma(p^2) \right \}  \  .
\end{align}
This prescription agrees with the usual definition $ \langle 0|
J^{a}_{\mu}|\rho_n^b \rangle = \delta^{ab}f_{n}\epsilon_{\mu} $ for
the vector meson decay constants. In our case,
\begin{align}
f^2_{n} = \frac{2 M_n^2} {g_5^2z_0^2  J_1^2(\gamma_{0,n})} \  .
\end{align}


\sect{Three-Point Function.} Consider  now the  trilinear term of
the action calculated on the $V(q,z)$ solution:
\begin{align}\label{action}
S_{\rm AdS}^{(3)}  &=
 -\frac{\epsilon_{abc}}{2g^2_5}\int d^4 x
 \int^{z_0}_{\epsilon}\frac{dz}{z}\,
 \left(\partial_{\mu}A_{\nu}^a\right)A^{\mu,b}A^{\nu,c} \ .
\end{align}
A na{\"\i}ve  variation gives   the   result  for   the 3-point
correlator $\langle J_a^{\alpha}(p_1)J_b^{\beta}(-p_2)
J_c^{\mu}(q)\rangle$ that contains the isotopic Levi-Civita tensor
$\epsilon_{abc}$,  the  dynamical   factor
\begin{align}
\label{Fsuv} W(p_1,p_2,q) \equiv \int^{z_0}_{\epsilon}
{\frac{dz}{z}}\, \frac{V(p_1,z)}{V(p_1,\epsilon)} \,
\frac{V(p_2,z)}{V(p_2,\epsilon)} \, \frac{V(q,z)}{V(q,\epsilon)} \
,
\end{align}
and  the tensor  structure
\begin{align}
T^{\alpha \beta \mu} =  \eta^{\alpha \mu }(q - p_1)^{\beta} -
\eta^{\beta \mu }(p_2+q)^{\alpha} + \eta^{\alpha\beta} (p_1 +
p_2)^{\mu}  \nonumber
\end{align}
familiar from  the QCD 3-gluon vertex amplitude.  Restoring the
transverse   projectors $\Pi^{\alpha \alpha'}(p_1)$, etc.
one   can   convert it   into
\begin{align}
\label{Tabm}
 {\cal T}^{\alpha\beta\mu} &=
\eta_{\alpha\beta}(p_1+p_2)_{\mu} + 2(\eta_{\alpha \mu}q_{\beta} -
\eta_{\beta \mu}q_{\alpha})    \ .
\end{align}

\noindent  For the  factors corresponding to the   hadronized
channels,  the Kneser-Sommerfeld expansion (\ref{KS})   gives
\begin{align}\label{bulkboundary}
&\frac{V(p,z)}{V(p,0)}  \equiv {\cal V}(p,z) = -g_5  \sum_{n =
1}^{\infty}\frac{ f_{n} \psi_n( z)}{p^2 - M^2_{n} } \ ,
\end{align}
where $p$  equals $p_1$ or $p_2$,   and
\begin{equation}\label{holographicwf}
\psi_n( z)   = \frac{\sqrt{2} }{z_0 J_1(\gamma_{0,n})}\, z J_1(M_{n}
z)
\end{equation}
is  the  ``$\psi$  wave function'' obeying the   same equation of
motion (\ref{Vequation}) as $V(p,z)$  (with $ p^2 = M^2_{n} $),
satisfying the b.c.
\begin{align}\label{bc1}
\psi_n(0) = 0  \  \  ,    \   \
\partial_z \psi_n(z_0) = 0  \  ,
\end{align}
and normalized according to
\begin{align}\label{normal}
\int^{z_0}_0~ \, \frac{dz}{z} \, |\psi_n(z)|^2 = 1  \ .
\end{align}
 One   remark  is in  order   here. 
Since the ``$\psi$  wave functions''   vanish   at   the  origin and  satisfy  
Neumann   b.c.    at   the IR   boundary,   it   is   impossible
to establish   a direct   analogy between $\psi_n  (z)$'s   and the 
bound state wave   functions in quantum   mechanics.
For   the   latter,  one  would  expect   that  they   vanish  at   the confinement   radius, 
while their   values  at   the origin
are  proportional  to the      coupling  constants $f_n$.

Taking a  spacelike   momentum   transfer, $ q^2 = - Q^2 $ for the
$V/V$ factor of  the EM   current   channel gives
\begin{equation}
{\cal J}(Q,z) = {Qz} \left  [ K_1(Qz) + I_1(Qz)
\frac{K_0(Qz_0)}{I_0(Qz_0)} \right  ] \   ,  
\end{equation}
 the non-normalizable   mode with Neumann b.c.  (see  also Ref.~\cite{Hong:2004sa}). 
This factor   can   also   be written as a  sum of monopole
contributions    from the infinite tower of vector mesons:
\begin{equation}
\label{Jmeson} {\cal J}(Q,z) =
 g_5  \sum_{m = 1}^{\infty}\frac{ f_{m}
\psi_m( z)}{ Q^2 + M^2_{m} } \ ,
\end{equation}
This decomposition,   discussed in  Ref.~ \cite{Hong:2004sa},
directly   follows from  Eq.~(\ref{bulkboundary}).
 Incorporating the representation for  the  bulk-boundary propagators
given above we  obtain
\begin{align}
\label{threepoint} & T(p^2_1, p^2_2, Q^2) = \sum_{n,k =
1}^{\infty}\frac{f_{n} f_{k}  F_{nk} (Q^2) }{\left(p_{1}^2 -
M^2_{n}\right)\left(p_{2}^2 - M^2_{k}\right)}  \  ,
\end{align}
where  $T(p^2_1, p^2_2, Q^2) =  W (p_1,p_2,q)/g_5^2$,  and
\begin{equation}
\label{ff} F_{nk}(Q^2)  = \int^{z_0}_{0} \frac{dz}{z} \,  {\cal J}
(Q,z) \, \psi_n ( z)  \,  \psi_k (z)
\end{equation}
correspond to    form factors for $n \to k$  transitions.
This  expression  was also  written  in  Ref.~\cite{Hong:2004sa}
for  form   factors considered    there.


\sect{Wave  functions.}
The formulas obtained above    using explicit properties of   the
Bessel  functions   in  the form of Kneser-Sommerfeld  expansions,
can    also  be derived   from the general formalism of Green's
functions. In  particular,   the Green's function for
Eq.~(\ref{Vequation}) can   be   written as
\begin{align}
G(p;z,z') = \sum^{\infty}_{n = 1}\frac{\psi_n(z)\psi_n(z')}{p^2 -
M^2_n} \ ,
\end{align}
where $ \psi_n(z) $'s are   the normalized wave functions
(\ref{holographicwf})  that  satisfy the Sturm-Liouville equation
(\ref{Vequation}) with \mbox{$ p^2 = M^2_n $} and Neumann b.c.
(\ref{bc1}). As  discussed in Ref.~\cite{Erlich:2005qh}, the
bulk-to-boundary propagator is related to the  Green's function by
\begin{align}\label{greensway}
{\cal V}(p,z') = - \left[ \frac{1}{z}\, \partial_z
G(p;z,z')\right]_{z = \epsilon \rightarrow 0} \ ,
\end{align}
and the two-point  function  $\Sigma(P^2)$ is   obtained from   the
Green's function using  Eqs.~(\ref{sigmap2}),(\ref{greensway})
\begin{align}
\label{sigmaG} \Sigma(P^2) =
\frac{1}{g^2_5}\left[\frac{1}{z'}\,\partial_{z'}\left[\frac{1}{z}\,
\partial_z G(p;z,z')\right]\right]_{z,z' = \epsilon \rightarrow 0} \
.
\end{align}
Accordingly, the coupling constants are  related   to  the $\psi$
wave  functions by
\begin{equation}
 f_n =\frac1{g_5} \,  \left  [ \frac1{z}\, \partial_z \psi_n (z) \right ]_{z=0}
\end{equation}
(cf.  \cite{Hong:2004sa,Erlich:2005qh}). In   view  of    this   relation, it
makes   sense to introduce ``$\phi$ wave functions''
\begin{equation}
 \phi_n(z) \equiv  \frac1{M_n z}\, \partial_z \psi_n (z) = \frac{\sqrt{2} }{z_0  J_1(\gamma_{0,n})}\,
J_0(M_{n} z)  \  ,
\end{equation}
which give   the couplings $g_5 f_n/M_n$ as   their values at   the
origin.   In   this   respect,  the  ``$\phi$ wave functions''  
are analogous   to   the
bound   state wave functions  in  quantum mechanics.    Moreover,  
these functions    satisfy   Dirichlet b.  c.
$\phi_n(z_0)=0$  and   are normalized   by
\begin{equation}
 \int_0^{z_0} dz \, z \,  | \phi_n(z)|^2   =   1 \  ,
\end{equation}
which strengthens   this  analogy. 
However,   the   elastic  form factors
$F_{nn}(Q^2)$ are   given by  the     integrals
\begin{align}
\label{form1} F_{nn}(Q^2) = \int^{z_0}_{0} \frac{dz }{z} \,  {\cal
J} (Q,z) \, |\psi_n (z) |^2
\end{align}
 involving
$\psi$ rather   than $\phi$ wave functions.
 In
fact,   due to the  basic equation (\ref{Vequation}), $\psi_n(z)$
wave functions    can   be expressed  in terms  of $\phi_n (z)$  as
\begin{equation}
 \psi_n (z)=- \frac{z}{M_n }\  \partial_z \phi_n (z) \   ,
\end{equation}
and we can  rewrite the   form   factor   integral as
\begin{align}
\label{form2} F_{nn}(Q^2) = &\int^{z_0}_{0} {dz }\, {z} \,  {\cal J}
(Q,z) \, |\phi_n (z) |^2  \\[7pt] \nonumber  &+ \frac{1}{M_n} \int^{z_0}_{0}{dz
}\, \phi_n (z) \, \psi_n (z) \,  {\partial_z  {\cal J} (Q,z) }  \, \
.
\end{align}
Note,   that the  nonnormalizable mode
\begin{equation}
\label{JD0} \frac1z \,  \partial_z  \, {\cal J} (Q,z) = - {Q^2 }
\biggl [ K_0(Qz) -  I_0(Qz) \,  \frac{K_0(Qz_0)}{ I_0(Qz_0)} \biggr
]
\end{equation}
corresponds to   equation whose   solutions are the   functions $
J_0(M_n z)$ satisfying Dirichlet   b.c.  at \mbox{$z=z_0$. }
 Expressing $\phi_n(z)$  in   terms  of $\partial_z \psi_n (z)$,
integrating $|\psi_n(z)|^2  $   by   parts  and   using  equation
(\ref{Vequation}) for $  {\cal J} (Q,z) $  gives
\begin{align}
\label{form2a} F_{nn}(Q^2) = &\int^{z_0}_{0} {dz }\, {z} \,  {\cal
J} (Q,z) \, |\phi_n
(z) |^2  \\[7pt] \nonumber  &- \frac{Q^2}{2M_n^2} \int^{z_0}_{0} \frac{dz }{z}  \,
  {\cal J} (Q,z) \,
|\psi_n (z)|^2   \ .
\end{align}
The  second   term contains the   original   integral for
$F_{nn}(Q^2)$, and   we  obtain
\begin{equation}
\label{formfin} F_{nn}(Q^2) = \frac1{1+{Q^2}/{2M_n^2}}
 \int^{z_0}_{0} {dz }\, {z} \,  {\cal J} (Q,z) \, |\phi_n
(z) |^2    \ .
\end{equation}
Notice,   that   the   normalizable  modes $\phi_n(z)$ in this
expression correspond  to Dirichlet b.c., while  the nonnormalizable
mode ${\cal J} (Q,z)$ was obtained using the Neumann ones.

Thus, we  managed  to   get  the   expression 
for $ F_{nn}(Q^2) $  form   factors  that contains $\phi$   instead   of $\psi$ 
wave   functions.   However,  it   contains  an extra
 factor   \mbox{$1/(1+Q^2/2M_n^2)$,}
which   brings   us  to   the  issue of 
different   form   factors of   the   $\rho$-meson   and kinematic 
  factors   associated   with them.


\sect{Form Factors.} Our result (\ref{ff})  contains only   one
function for   each $n \to k$ transition, in   particular $F_{nn}
(Q^2)$ in the    diagonal  case.    However, the general expression
for the EM vertex of a spin-1 particle of   mass $M$ can   be
written (assuming $P$- and $T$-invariance) in terms of three form
factors (see, e.g., \cite{Arnold:1979cg}, our $G_2$ is theirs
$G_2-G_1$):
\begin{align}\label{rhoFF}
\langle \rho^{+}(p_2,\epsilon')& |J^{\mu}_{\rm EM}
(0)|\rho^{+}(p_1,\epsilon) \rangle \\[4pt] \nonumber = -
\epsilon'_{\beta} \epsilon_{\alpha} \bigl[& \ \eta^{ \alpha \beta}
(p_1^{\mu}  + p_2^{\mu} )\, G_1(Q^2) \\[3pt]  \nonumber +
 &(\eta^{\mu \alpha}q^{\beta} - \eta^{\mu \beta}q^{\alpha} )
(G_1(Q^2) + {G}_2(Q^2) )  \\  \nonumber - &\frac{1}{M^2} q^{\alpha}
q^{\beta} (p_1^{\mu}+p_2^\mu )\,  G_3(Q^2) \ \bigr]  \  . \nonumber
\end{align}
Comparing the tensor structure of this expression with (\ref{Tabm}),
we conclude that H-model predicts $G_1^{(n)}(Q^2)=G_2^{(n)}(Q^2) =
F_{nn}(Q^2)$, and $G_3^{(n)} (Q^2) =0$ for form  factors
$G_i^{(n)}(Q^2)$ of $n^{\rm th}$ bound state.
 It was   argued
(see \cite{Hong:2004sa})  that  this   is a  general
feature of AdS/QCD   models  for    the $\rho$-meson
form   factors.  
Since  ${\cal J}
(Q=0,z) =1$,  the diagonal form factors $F_{nn}(Q^2)$  in the
H-model are normalized to unity, while the nondiagonal  ones vanish
for $Q^2 =0$ (the functions $\psi_n (z)$  
are  orthonormal on $[0,z_0]$).

The form factors $G_i $ are related to electric $G_C$, magnetic
$G_M$ and quadrupole $G_Q$ form factors  by
\begin{eqnarray}
G_C &=& G_1 + \frac{Q^2}{6M^2} \, G_Q  \ , \ \ G_M  = G_1 + G_2 \ ,
\nonumber \\
G_Q  &=&  \left(1+\frac{Q^2}{4M^2}\right )G_3  - G_2   \ .
\end{eqnarray}
For these form factors, H-model gives
\begin{align}
&G_Q ^{(n)}(Q^2) = - F_{nn}(Q^2) \ , \ \ G_M^{(n)} (Q^2) = 2
F_{nn}(Q^2)\ , \\[7pt] \nonumber &G_C^{(n)} (Q^2) = \left  (1-
\frac{Q^2}{6M^2} \right )F_{nn}(Q^2) \ .
\end{align}
For $Q^2=0$, it correctly reproduces the unit electric charge of the
meson, and ``predicts'' \mbox{$\mu \equiv G_M(0) = 2$} for the
magnetic moment and \mbox{$D\equiv G_Q(0)/M^2= -1/M^2 $} for the
quadrupole  moment, which are just the canonical values for a
pointlike vector particle \cite{Brodsky:1992px}.

Another interesting combination of form factors
\begin{equation}
{\cal F} (Q^2) = G_1(Q^2) + \frac{Q^2}{2M^2}\, G_2(Q^2) - \left (
\frac{Q^2}{2M^2} \right )^2 \,G_3(Q^2)
\end{equation}
appears if one  takes the ``+++'' component   of the 3-point
correlator (obtained, e.g.,  by convoluting it with  $n_\alpha
n_\beta n_\mu$, where $n^2=0, (np_1)=1, (nq)=0$
\cite{Radyushkin:2006iz}). The H-model result (\ref{formfin})  for
${\cal F} (Q^2)$ is particularly simple:
\begin{equation}
\label{calF} {\cal F}_{nn}(Q^2) =
 \int^{z_0}_{0} {dz }\, {z} \,  {\cal J} (Q,z) \, |\phi_n
(z) |^2    \   .
\end{equation}
 Thus,  it   is  the  form   factors  ${\cal F}_{nn}(Q^2)$  that  are   the  most   direct
analogues of diagonal   bound state   form   factors in  quantum   mechanics.


\sect{Low-$Q^2$ behavior.} Our expression   for   ${\cal
F}_{nn}(Q^2)$ is close to   that  proposed for a  generic meson form
factor in   the holographic model   of   Ref.~\cite{Brodsky:2006uq}.
There,  the   authors   used ${\cal K} (Qz)  \equiv Qz K_1(Qz)$ as
the $q$-channel   factor.  Indeed, the    difference  between 
${\cal J} (Q,z)$   and ${\cal K} (Qz)$ is   exponentially  small
when  $Qz_0 \gg 1 $,   but the two   functions    radically differ
in   the region of  small $Q^2$,   where the function ${\cal K}
(Qz)$  displays  the  logarithmic branch singularity
\begin{equation}
 {\cal K} (Qz) = 1- \frac{ z^2 Q^2} {4} \Biggl  [ 1- 2 \gamma_E
- \ln (Q^2 z^2 /4) \Biggr  ]   +  {\cal O} (Q^4)  \  ,
\end{equation}
that leads   to   incorrect infinite   slope  at $Q^2=0$. To implant
the AdS/QCD information   about  the  hadron   spectrum in the
$q$-channel  one  should use ${\cal J}(Q,z)$  that corresponds to
a tower  of    bound   states in the $q$-channel. The lowest
singularity in this   case   is   located at \mbox{$Q^2=-M_1^2$}.
Since    it   is   separated   by   a finite    gap from    zero,
the  form   factor   slopes  at $Q^2=0$ are   finite.

To  analyze  the form factor behavior in  the $Qz_0 \ll 1$ limit, we
expand
\begin{equation}
\label{j} \left. {\cal J}(Q,z)  \right |_{Qz_0 \ll 1} = 1- \frac{z^2
Q^2}{4}\left [1-  \ln \frac{z^2}{z_0^2} \right ] + {\cal O} (Q^4)  \
.
\end{equation}
As   expected, the   result  is   analytic in $Q^2$. For  the
lowest transition (i.e.,   for   the $\rho$-meson form    factor),
explicit  numbers are  as follows:
\begin{equation}
 \label{calf11num}
  {\cal F}_{11}(Q^2)  \approx 1- 0.692  \, \frac{Q^2}{M^2} +
0.633  \, \frac{Q^4}{M^4} + {\cal O} (Q^6)   \ ,
\end{equation}
where  $M=M_1=m_\rho$. Another  small-$Q^2$  expansion
\begin{equation}
 \label{f11num}
  F_{11}(Q^2)  \approx 1- 1.192 \,  \frac{Q^2}{M^2} + 1.229 \,  \frac{Q^4}{M^4}+ {\cal O} (Q^6)   \ ,
\end{equation}
can be  either calculated from the original expression (\ref{form1})
involving $\psi$-functions or by dividing  ${\cal F}_{11}(Q^2) $ by
$(1+Q^2/2M^2)$.   The  latter  approach  easily  explains  the
difference in  slopes of these two   form   factors at  $Q^2=0$.
Finally,  for   the  electric   form   factor,  we obtain
\begin{equation}
 \label{Gcnum}
  G_{C}^{(1)} (Q^2)  \approx 1- 1.359 \,\frac{Q^2}{M^2} +1.428\,
\frac{Q^4}{M^4}  + {\cal O} (Q^6)   \  .
\end{equation}
For  the  electric radius of    the $\rho$-meson   this  gives
\begin{equation}
 \langle r_\rho^2 \rangle _C = 0.53 \, {\rm fm}^2 \   ,
\end{equation}
the value  that is very   close  to  the recent  result
\mbox{(0.54\, fm$^2$)}   obtained within the Dyson-Schwinger
equations  (DSE) approach \cite{Bhagwat:2006pu}. Lattice gauge
calculations  \cite{Lasscock:2006nh} indicate  a similar     value
in   the $m_\pi^2 \to 0$ limit.


\sect{Vector  meson  dominance patterns.} Numerically,  the result
$1.359/M^2$   for  the  slope  of  $G_C^{(1)}(Q^2)$ is larger   than
the simple VMD expectation $1/M^2$. In   fact,  a  part  of this larger
value   is   due   to   the factor
$(1-Q^2/6M^2)$ relating $G_C^{(1)}(Q^2)$ and $F_{11} (Q^2)$,
which is   kinematic  to some extent. The $F_{11} (Q^2)$
form factor,  however,  can   be   written in   the   generalized  VMD
representation 
 (cf. \cite{Hong:2004sa})
\begin{equation}\label{Gm11}
{ F}_{11}(Q^2) =  \sum_{m=1}^{\infty} \frac{  F_{m,11} }{ 1+ Q^2/
M_m^2}  \ ,
\end{equation}
with  the  coefficients $F_{m,11}$  given   by  the overlap
integrals
\begin{equation}
\label{fm11}  F_{m,11} ={4 } \int^{1}_{0} {d \xi} \, {\xi^2} \,
\frac{ J_1 (\gamma_{0,m} \xi) \,  J_1^2 (\gamma_{0,1}
\xi)}{\gamma_{0,m}J_1^2(\gamma_{0,m}) J_1^2(\gamma_{0,1})}  \ ,
\end{equation}
apparently   having a purely dynamical  origin. The coefficients
$F_{m,11}$   satisfy  the   sum   rule
\begin{equation}
 \label{sumGm11}
   \sum_{m=1}^{\infty}
 {  F_{m,11} }  =1
 \end{equation}
that    provides correct normalization $F_{11}(Q^2=0)=1$.
Numerically,   the  unity   value  of    the    form    factor
$F_{11}(Q^2)$  for  $Q^2=0$ is  dominated by the   first bound state
that  gives $1.237$. The   second bound   state makes a sizable
correction by $-0.239$, while adding a   small 0.002 contribution
from the  third bound state   fine-tunes 1   beyond the $10^{-3}$
accuracy. Contributions from   higher   bound   states to  the  form
factor  normalization are   negligible at   this   precision.

The slope of  $F_{11}(Q^2)$  at  $Q^2=0$ is   given   by   the sum
of $F_{m,11} /M_m^2$ coefficients.   Now,     the  dominance   of
the first bound state is   even   more   pronounced: the  $Q^2$
coefficient $1.192/M^2$ in  Eq.~(\ref{f11num}) is   basically
contributed by the   first bound   state   that  gives $1.237/M^2$,
with small $-0.045/M^2$ correction  from the   second bound   state.
Other   resonances are   not   visible at  the  three-digit
precision.

Thus,  for   small $Q^2$,   H-model  predicts  a  rather peculiar
pattern of   VMD for  $F_{11}(Q^2)$ 
 (observed originally
in   Ref.~\cite{Hong:2004sa} for   a  form   factor   considered there):
 strong dominance  of   the
first $q$-channel bound state, whose   coupling $F_{1,11}$  exceeds
1, with   the   second resonance (having the negative coupling
$F_{2,11}$) compensating   this   excess.

Similarly, the   ${\cal F}_{11}  (Q^2)$  form   factor  has the
generalized VMD representation with   coefficients ${\cal
F}_{m,11}$ given   by    the  overlap integrals
\begin{equation}
\label{calfm11}  {\cal F}_{m,11} ={4 } \int^{1}_{0} {d \xi} \,
{\xi^2} \, \frac{ J_1 (\gamma_{0,m} \xi) \,  J_0^2 (\gamma_{0,1}
\xi)}{\gamma_{0,m}J_1^2(\gamma_{0,m}) J_1^2(\gamma_{0,1})}  \  .
\end{equation}
Now, $  {\cal F}_{1,11} \approx  0.619$, $ {\cal F}_{2,11} \approx
0.391$, $  {\cal F}_{3,11} \approx  -0.012$, $  {\cal F}_{4,11}
\approx  0.002$,  etc.   In  this   case also,   the   value  of the
${\cal F}_{11}  (Q^2)$  form factor for  $Q^2=0$ is dominated by the
first two bound states. For the slope of   the form factor at
$Q^2=0$, the  dominance of the first bound state is again more
pronounced: the  $Q^2$ coefficient  $0.692/M^2$ in
Eq.~(\ref{calf11num}) is   basically contributed by the   first
bound   state   that  gives $0.619/M^2$, with a small $0.074/M^2$
correction  from the   second bound   state   and a tiny $-0.001/M^2$
contribution from the third one.

Thus, for ${\cal F}_{11}  (Q^2)$, H-model gives again a
two-resonance dominance pattern, with the coupling  $ {\cal
F}_{2,11}$ of   the  second   resonance being now  just  somewhat
smaller than  the  coupling   $ {\cal F}_{1,11}$ of   the  first 
resonance,   both   being  positive. The   relation between the two
VMD   patterns follows   from Eq.~(\ref{formfin}):
\begin{equation}
{ F}_{m,11}  =  \frac{{\cal F}_{m,11}}{1- {M_m^2}/{2 M_1^2}}  \  .
\end{equation}
In   particular,   it  gives ${ F}_{1,11}  =  2  \, {\cal
F}_{1,11}$, and negative   sign for ${ F}_{2,11}$. It  also
determines  that if  higher  coefficients  ${\cal F}_{m,11}$ are
small  then ${ F}_{m,11}$'s   are  even   smaller.


\sect{Large-$Q^2$ behavior.} 
Eq.~(\ref{formfin})   tells  us that
asymptotically   ${F}_{11}  (Q^2)$ is  suppressed by a  power of
$1/Q^2$  compared to  ${\cal F}_{11}  (Q^2)$, which is  known to
behave like  $1/Q^2$   for  large $Q^2$
\cite{Brodsky:2006uq,Radyushkin:2006iz}. The absence   of  $1/Q^2$
term in    the asymptotic   expansion for  ${F}_{11}  (Q^2)$ means
that the  coefficients $F_{m,11}$   defined in  Eq.~(\ref{Gm11})
satisfy the ``superconvergence'' relation
\begin{equation}
\sum_{m=1}^\infty M_m^2 F_{m,11} =0  
\end{equation}
reflecting a  ``conspiracy'' \cite{Hong:2004sa} between   the poles.
 Writing $ M_m^2 F_{m,11} \equiv A_m M^2$,  we  obtain that $A_1
\approx  1.237$, $A_2   \approx - 1.261$, $A_3 \approx  0.027$
 (our   results for  the  ratios $A_2/A_1$, $A_3/A_1$  agree   with  
the   calculation of  Ref.~\cite{Hong:2004sa}).
Again, the sum   rule is practically saturated by the  first  two
bound    states,  which give contributions that   are close in
magnitude  but  opposite in   sign.

In   case of  ${\cal F} (Q^2)$, the two lowest bound states both
give  positive ${\cal O} (1/Q^2)$ contributions at large $Q^2$. In
Ref.~\cite{Radyushkin:2006iz}, it   was shown that the  asymptotic
normalization  of  ${\cal F}_{11} (Q^2)$ exceeds   the VMD
expectation $M_1^2/Q^2$   by a factor of 2.566. We can infer this
normalization  from the values of the   coefficients ${\cal
F}_{m,11}$  defined in  Eq.~(\ref{calfm11}). Writing $ M_m^2 {\cal
F}_{m,11} \equiv {\cal A}_m M_1^2$,  we  obtain that ${\cal A}_1
\approx  0.619$, ${\cal A}_2   \approx 2.061$, ${\cal A}_3 \approx
-0.150$, ${\cal A}_4 \approx  0.054$. Note,  that  the total  result
is dominated by the  second  bound    state,  which  is responsible
for  about 80\% of   the   value.  The   lowest  bound  state
contributes only   about 25\%,   while   the  higher   states give
just   small   corrections.

It  is worth    noting  that the  large-$Q^2$ behavior   of ${\cal
F}_{11} (Q^2)$ is   determined by  the  large-$Qz_0$ form of ${\cal
J}(Q,z)$:  it   can   be (and was)  calculated using ${\cal K}
(Qz)$,   the free-field   version of  ${\cal J}(Q,z)$.   As  a
result,  the    value  of  the asymptotic coefficient (2.566 in
case of ${\cal F}_{11} (Q^2)$) is  settled  by the   sum  rule
\begin{equation}
\sum_{m=1}^\infty M_m^2 {\cal F}_{m,11} =  |\phi_1(0)|^2
\int_0^\infty d\chi \, \chi^2 \,  K_1 (\chi) = 2 \, |\phi_1(0)|^2
\end{equation}
that should be satisfied by any  set of coefficients  ${\cal
F}_{m,11}$. A particular distribution of ``2.566'' among the bound
states is  governed by  the  specific $q$-channel dynamics (in our
case, by the choice of the Neumann b.c. for ${\cal J}(Q,z)$  at
$z=z_0$). Thus, in the dynamics described by ${\cal J}(Q,z)$, the
large value of the asymptotic coefficient is explained by large
contribution due to the second bound state.

 It    was  shown in   Ref.~\cite{Radyushkin:2006iz} that 
the asymptotic $1/Q^2$  
behavior for  ${\cal F}_{11}  (Q^2)$ is   established only    for 
$Q^2 \sim 10\,  {\rm GeV}^2$,  and   one   may 
question the  applicability of  the H-model  for  such   large
$Q^2$.  The   discussion   of   this  problem,  however,   is  beyond    the
scope  of   the   present paper.


\sect{Summary.} In this letter, we described the formalism that
allows to study form factors of vector mesons in the holographic QCD
model of Refs.~\cite{Erlich:2006hq,Erlich:2005qh,DaRold:2005zs}
(H-model). An essential ingredient of our approach is  a  systematic use
of the Kneser-Sommerfeld representation that explicitly displays the
poles of two- and three-point functions and describes the structure
of the corresponding bound states by eigenfunctions of the 5D
equation of motion, the ``$\psi$ wave functions''. These functions
vanish at $z=0$ and satisfy Neumann b.c. at $z=z_0$, which prevents
a direct analogy with bound state wave functions in quantum
mechanics. To this end, we introduced  an alternative description in
terms of ``$\phi$ wave functions'' that   satisfy Dirichlet b.c. at
$z=z_0$ and  have finite  values at $z=0$ which determine bound
state couplings $g_5 f_n/M_n$.
 Thus,   the $\phi$ wave   functions 
have   the   properties  necessary   for   the 
light-cone interpretation proposed in  Ref.~\cite{Brodsky:2006uq}
and      discussed also in   Ref.\cite{Radyushkin:2006iz}.

Analyzing the three-point function, we derived expressions for bound
state form factors both in terms of $\psi$ and $\phi$ wave
functions, and obtained   specific predictions for  form factor
behavior at small and large values of the invariant momentum
transfer $Q^2$.  In particular, we calculated the  electric radius
of the $\rho$ meson, and obtained the value $\langle r_\rho^2
\rangle _C = 0.53 \, {\rm fm}^2$ that practically coincides with the
recent result \cite{Bhagwat:2006pu} obtained within the DSE
approach. Our result is also consistent with the $m_\pi^2 \to 0$
extrapolation of the recent lattice  gauge calculation
\cite{Lasscock:2006nh}.

We derived a generalized VMD representation 
 both   for the $F_{11}(Q^2)$  form    factor
(the  expression for   which coincides  with  a   model
 $\rho$-meson   form   factor 
considered in Ref.~\cite{Hong:2004sa})   and  for  the
${\cal F}_{11}(Q^2)$  form   factor  introduced in    the   present   paper, 
and demonstrated that H-model predicts a very specific VMD pattern, in
which these form factors are essentially given by contributions due to the
first two bound states in the $Q^2$-channel, with the higher bound
states playing a negligible role. We showed that, while the form
factor slopes at $Q^2=0$ in this picture are dominated by the first
bound state, the second bound state plays a crucial role  in the
large-$Q^2$ asymptotic limit. In particular, it provides the bulk
part of the negative  contribution necessary to cancel the
na{\"\i}ve VMD $1/Q^2$ asymptotics for the $F_{11}(Q^2)$ form factor
(corresponding to the overlap integral involving the $\psi$
functions), and it dominates the asymptotic $1/Q^2$ behavior of the
${\cal F} (Q^2)$ form factor (given by the overlap of the $\phi$
functions).

A possible future application of our approach is the analysis of
bound state form factors in the model of Ref.~\cite{Karch:2006pv} in
which the hard-wall boundary conditions at the $z=z_0$ IR boundary
are substituted by an oscillator-type potential.  This model
provides the $ M^2_n \sim n \Lambda^2$ asymptotic behavior of the
spectrum of highly excited mesons,  which  is   more consistent with
the semiclassical limit of QCD \cite{Shifman:2005zn} than the $M^2_n
\sim n^2 \Lambda^2$ result of the H-model.

\sect{Acknowledgments.} H.G. would like to thank J. Erlich for
illuminating discussions, A.~W. Thomas for valuable comments and
support, J.~L. Goity and R.~J. Crewther for  stimulating
conversations, J.~P. Draayer for support at Louisiana State
University, and G.~S.  Pogosyan and S.~I. Vinitsky for support at
JINR, Dubna. A.R. thanks J.~J. Dudek for attracting attention to
Ref.~\cite{Lasscock:2006nh}.

Notice: Authored by Jefferson Science Associates, LLC under
 U.S. DOE Contract No. DE-AC05-06OR23177. 
The U.S. Government retains a non-exclusive, paid-up, irrevocable, 
world-wide license to publish or reproduce this manuscript for U.S. Government purposes.

\end{document}